\documentclass[submission,copyright,creativecommons]{eptcs}
\providecommand{\event}{AFL 2023} 

\usepackage{latexsym,amssymb,amsmath}
\usepackage{graphics}
\newcommand{\eps}{\varepsilon}
\def\lra{\longrightarrow}
\def\ra{\rightarrow}
\def\bbbn{{\rm I\!N}}
\def\F{{\rm I\!F}}
\def\l{\langle}
\def\r{\rangle}

\def\bbbz{{\mathchoice {\hbox{$\sf\textstyle Z\kern-0.4em Z$}}
		{\hbox{$\sf\textstyle Z\kern-0.4em Z$}}
		{\hbox{$\sf\scriptstyle Z\kern-0.3em Z$}}
		{\hbox{$\sf\scriptscriptstyle Z\kern-0.2em Z$}}}}

\newcommand{\Lra}{\Longrightarrow}
\newcommand{\Ra}{\Rightarrow}

\newlength{\cwidth}

\newcommand{\N}{{\bbbn}}
\newcommand{\bean}{\begin{eqnarray*}}
	\newcommand{\eean}{\end{eqnarray*}}
\newtheorem{teor}{Theorem}
\newtheorem{lema}{Lemma}
\newtheorem{coro}{Corollary}
\newtheorem{prop}{Proposition}

\title{On the Degree of Extension of Some Models Defining Non-Regular Languages\thanks{This work was performed through the Core Program within the National Research, Development and Innovation Plan 2022-2027, carried out with the support of MRID, project no. 23020101(SIA-PRO), contract no 7N/2022, and project no. 23020301(SAFE-MAPS), contract no 7N/2022.}}
\author{Victor Mitrana
\institute{Department of Information Systems,\\ Universidad Polit\' ecnica de Madrid\\
	Calle Alan Turing s/n (Carretera de Valencia Km 7),\\ 28031 Madrid, Spain\\
	and\\
National Institute of R\&D for Biological Sciences, \\ 296 Independenței Bd.,
060031, Bucharest. Romania
\email{victor.mitrana@upm.es}}
\and \qquad Mihaela P\u aun
\institute{National Institute for R\&D for Biological Sciences \\
296 Independenței Bd., 060031, Bucharest, Romania
	\email{mihaela.paun@incdsb.ro}}}

\def\titlerunning{On the Degree of Extension}
\def\authorrunning{V. Mitrana, M. P\u aun}
\begin{document}
\maketitle

\begin{abstract}
This work is a survey of the main results reported for the degree of extension of two models defining non-regular languages, namely the context-free grammar and the extended automaton over groups. More precisely, we recall the main results regarding the degree on non-regularity of a context-free grammar as well as the degree of extension of finite automata over groups. Finally, we consider a similar measure for the finite automata with translucent letters and present some preliminary results. This measure could be considered for many mechanisms that extend a less expressive one.
\end{abstract}

\section{Introduction}

Language defining models play a central role in formal language theory, and in
theoretical computer science. There have been defined very many such models with various motivations depending on the specific problems to be solved. In this work, we restrict ourselves to the most well-known devices: Chomsky generative grammars and finite automata.
Regular languages are classically represented
by: regular or right-linear grammars, many variants of finite automata, regular expressions, logical or algebraic formalisms. Due to their limited expressiveness, some of these models have been extended to more complex models such that the old model is just a particular case of the extended one. For instance, context-free grammars are natural extensions of regular or right-linear grammars, finite automata with valences \cite{fernau}, jumping automata \cite{jumping}, automata with translucent letters \cite{translucent} are extensions of finite automata able to accept non-regular languages, etc.
In their turn, context-free languages are classically represented by: context-free grammars, pushdown automata, logical and algebraic formalisms. Mainly, by the same reason as above, there have been proposed various extensions like context-sensitive grammars, grammars with regulated rewriting \cite{regulated}, etc.
 
In this work, we recall a measure for evaluating the degree of extension of a two such models, namely the context-free grammar as an extension of regular grammar, and the extended finite automaton over groups as an extension of the finite automaton. A similar measure is also considered for finite automata with translucent letters.
Roughly speaking, this measure is defined as follows:\\
(i) by counting the maximal number of non-regular rules used in a derivation \cite{jcss},\\
(ii) by evaluating the group memory used by the extended automata over groups \cite{jlac}.\\
(iii) by counting the number of jumping moves used by a finite automata with translucent letters.\\
As far as the first measure is concerned, it is worth noting that similar investigations have been reported from the time of introducing the classes of regular and context-free languages. Here are several results giving sufficient conditions for a context-free grammar and a context-sensitive grammar to generate a regular and context-free language, respectively:
\begin{itemize}
	\item Each context-free grammar that is not self-embedding generates a regular language \cite{chomsky}.
	\item An arbitrary grammar in which no terminal is used as context and every rule generates at least one terminal, generates a context-free language \cite{GG}.
	\item An arbitrary grammar generates a context-free language if the
	left side of every rule contains only one nonterminal, with terminal words as
	the only context \cite{Book}.
	\item If every rule of an arbitrary
	grammar has as left context a word of terminal symbols at least as long as
	the right context, then the language generated is context-free \cite{Book}.
	\item A grammar which has a partial ordering  on its symbols, such that in every rule of the grammar every symbol on the left side is ``smaller'' than some symbol on the right generates a context-free language \cite{Hibbard}.	
	\item In a grammar, the sets of terminal words generated by ``one-way'' and ``two-ways'' derivations are context-free \cite{Matthews1,Matthews2,Matthews3} and \cite{Evey}.
	\item An arbitrary grammar such that in each of its non-context-free rules, the right side contains a word of terminals longer than any terminal word appearing between
	two nonterminals in the left side, generates a context-free language \cite{baker}.
\end{itemize}
As one can see, some of the above conditions can be immediately checked, namely by examining the rules.
In many cases, the complexity of a device generating a language is a function with nonnegative integer values: rational index \cite{Boasson}, initial index \cite{Gabarro2}, index of a context-free grammar \cite{Brainerd}, height of derivational trees \cite{Culik}, etc. Similar approaches have been reported in \cite{jcss} for context-free and context-sensitive grammars, and \cite{jlac} for extended automata over groups. In the sequel, we survey the most important results of these papers.

\section{Preliminaries}
We assume the reader is familiar with the basic definitions and 
concepts in formal language and automata theory and combinatorial algebra such as monoids and groups, presentations and generating sets, etc. For further details,
we refer to \cite{handbook} (for formal languages and automata theory), and \cite{Lyndon,Rotman} (for combinatorial algebra).

We denote by $\mathbb{N}$ the set of nonnegative integers. An alphabet is a finite set of letters or symbols. For a set $A$ we denote by
$card(A)$ the cardinality of $A$. For a finite set $V$, called alphabet, we denote by $(V^*,\cdot,\eps)$ the free monoid generated by $V$  under the
operation of concatenation with the neutral element $\eps$. The elements of $V^*$ are called words and $\eps$ is the empty word. For a word $x\in V^*$ we denote by $alph(x)$ the smallest subset of $V$ such that $x\in alph^*(x)$. Given a set $A$, we denote by ${\cal P}_f(A)$ the family of all
finite subsets of $A$.
The free semigroup generated by $V$ with concatenation is denoted by 
$V^+$. The length of $x \in V^*$ is denoted by $|x|$,
$|x|_a$ is the number of occurrences of $a$ in $x$, whereas 
$|x|_B$ is the number of occurrences of symbols $B\subseteq V$ in $x$.
For a word $w=a_1a_2\dots a_n$, $n\ge 1$, $a_i\in V$ for all $1\le i\le n$, we write $\widetilde{w}=a_n\dots a_2a_1$.

By regular grammar we mean a grammar that is right-linear, hence a regular rule should be understood as a right-linear rule of one of the forms $A\ra wB$, and $A\ra w$, with $A,B$ being nonterminals and $w$ being a word of terminals, possibly the empty word.
In what follows we also use the regular expressions for defining regular languages.
A context-free grammar $G=(N,T,S,P)$ is a \emph{reduced} grammar if for any $X\in N$ we have the derivations
$S\Lra^* \alpha X\beta$, for some $\alpha,\beta\in (N\cup T)^*$ ($X$ is said to be {\it accessible}), and $X \Lra^* u$, with $u\in T^*$ ($X$ is said to be {\it co-accessible}).
A context-free grammar is \emph{proper} if it has no $\lambda$-production (i.e. $X\ra\lambda$, $X\in N$) and no chain-production
(i.e., $X\ra  Y$, $X, Y \in N$). It is known that for every context-free grammar (which does not generate $\lambda$)
there exists an equivalent proper and reduced context-free grammar.

For an arbitrary grammar $G=(N,T,S,P)$ we denote by
\begin{itemize}
	\item $G(A)=(N,T,A,P)$, $A\in N$ the grammar in which the axiom $S$ was replaced by
	another nonterminal, $A$.
	\item $G_{reg}=(N,T,S,P_{reg})$ the grammar obtained from $G$ by considering the set
	$P_{reg} \subseteq P$ of regular productions of $P$ only.
	\item $G_{cf}=(N,T,S,P_{cf})$ the grammar obtained from $G$ by considering the set
	$P_{cf} \subseteq P$ of context-free productions of $P$ only.
\end{itemize}
We denote by $REG$ and $CF$ the class of regular and context-free languages, respectively.

A finite \emph{multiset} over a finite set $A$ is a mapping $\sigma:A\lra \bbbn$; $\sigma(a)$
expresses the number of copies of $a\in A$ in the multiset $\sigma$.
In what follows, a multiset containing the elements $b_1,b_2,\dots,b_r$, any element possibly being repeated one or more times in the sequence, will be denoted by $\l b_1,b_2,\dots, b_r\r$.
Each multiset $\sigma$ over a set $A$ of cardinality $n$ may also be viewed as an array of size $n$ with non-negative entries.

For two functions $f,g:\mathbb{N}\lra \mathbb{N}$ we say that $f(n)\in\mathcal{O}(g(n))$ iff there is a constant $c>0$ and $n_0\ge 1$ such that $f(n)\le cg(n)$ for all
$n\ge n_0$. Equivalently, $f(n)\in\mathcal{O}(g(n))$ iff $\displaystyle{\lim_{n\to\infty}\sup \frac{f(n)}{g(n)} < \infty}$. Following \cite{Balcazar}, we say that $f(n)\in\Omega(g(n))$ iff $\displaystyle{\lim_{n\to\infty}\sup \frac{f(n)}{g(n)} > 0}$.
Furthermore, we say that $f(n)\in o(g(n))$ iff $\displaystyle{\lim_{n\to\infty}\frac{f(n)}{g(n)}=0}$.

\section{The degree of non-regularity}

Given a context-free grammar $G=(N,T,S,P)$ a derivation step in $G$ by using the rule $r\in P$ is denoted by $\Ra_r$. For a derivation
in $G$
$$D=(S\Ra_{r_1}w_1\Ra_{r_2}w_2\cdots\Ra_{r_m}w_m=w),$$
where $w\in T^*$ and $r_i\in P$ for $1\leq i\leq m$, we define
the degree of non-regularity of $w$ with respect to $D$ by
$$dnreg_G(w,D)=\mbox{card}\{i\mid r_i\notin P_{reg}, 1\leq i\leq m\}.$$
Less formally, $dnreg_G(w,D)$ is the number of non-regular rules applied in the derivation $D$ of $w$ in the grammar $G$.
The degree of non-regularity of a terminal word $w$ with respect to
the grammar $G$ is
$$dnreg_G(w)=\left\{
\begin{array}{ll}
min\{dnreg_G(w,D)\mid D\mbox{ is a derivation of }w\mbox{ in }G\},\\
0, \mbox{ if } w\notin L(G).
\end{array}\right.$$
In other words, the degree of non-regularity
of a word with respect to a grammar is computed by taking into consideration the
``least non-regular derivation'' if there is one.

The degree of non-regularity of a context-free grammar $G$ as above is a mapping from $\mathbb{N}$ to $\mathbb{N}$ defined by
$$ dnreg_G(n)= max\{dnreg_G(w)\mid |w|=n, w\in T^+\}.$$
As one can see, the most ``non-regular'' word of each length is considered.

For a function $f:\mathbb N\ra\mathbb N$ we now define the complexity class
\bean
DNREG (f(n))&=&\{L\mid L=L(G) \mbox{ for some context-free grammar $G$ and } \\
&& dnreg_G(n)\in \mathcal O(f(n))\}.
\eean

Otherwise stated, a language has the degree of non-regularity $f(n)$ if and only if it belongs to $DNREG (f(n))$.

In the sequel we recall the main results about 
the degree on non-regularity. A simple remark turns out to be useful.
If $G$ is an arbitrary context-free grammar and $G_1$ is the reduced grammar obtained from $G$, $dnreg_G(n)=dnreg_{G_1}(n)$ holds for all $n$, because none of the removed productions contributes in any derivation of a terminal word in $G$.
By several considerations, a similar situation holds if $G$ is not proper and $G_1$ is the proper grammar obtained from $G$. Therefore, the context-free grammars considered in the sequel are reduced and proper.

A context-free grammar is said to be in \emph{quasi normal form} if all its
rules of are of the following forms:
\bean
&(i)& A\ra a, \mbox{ where $a$ is a terminal},\\
&(ii)& A\ra aB, \mbox{ where $a$ is a terminal and $B$ is a nonterminal},\\
&(iii)& A\ra \alpha, \mbox{ where $\alpha$ is a word of nonterminals of length at least $2$}.
\eean

\begin{prop}\label{normal form}
	For every context-free grammar $G$ there exists an equivalent context-free grammar $G'$ in quasi normal form such that $dnreg_G(n)=dnreg_{G'}(n)$ for all $n$.\\		
\end{prop}

If the length of $\alpha$ is exactly $2$ in every rule $A\ra\alpha$ of a grammar in quasi normal
form, we say that the grammar is in {\it quasi Chomsky normal form}.
If we have a grammar in quasi normal form, each rule $A\ra\alpha$, with 
$|\alpha|\ge 3$ can be replaced by 
a sequence of rules with the right-hand side of length $2$. 
Hence, each grammar in quasi normal can be replaced by an equivalent 
grammar in quasi Chomsky normal form at a price of a constant number of times higher degree of non-regularity.

Let $G$ be a context-free grammar and $c$ be a positive integer; we define the language
$$L(G,\le c)=\{w\in L(G)\mid dnreg_G(w)\le c\}.$$
Clearly, if $dnreg_G(n)\le c$ for any $n\ge 1$, then $L(G,\le c)=L(G)$ holds.

\begin{teor} $DNREG(1)=REG$.
	A language generated by a context-free grammar is finitely-non-regular if and only if it is regular.
\end{teor}

\begin{teor}\label{egal_constanta}
	For any given context-free grammar $G$ and a positive integer $c$, one can algorithmically decide whether $dnreg_G(n)\le c$.
\end{teor}

It is worth mentioning that $dnreg_G(n)\le c$, for a context-free grammar $G$ and a positive integer $c$, implies that $L(G)$ is regular. However, if
$dnreg_G(n)> c$, then nothing can be said about the regularity of $L(G)$. Even more, if $L(G)$ is regular, it does not generally follow that $dnreg_G(n)\in \mathcal{O}(1)$.

\begin{teor}
	Given an unambiguous context-free grammar $G$, one can algorithmically decide whether
	$dnreg_G(n)\in \mathcal O(1)$.
\end{teor}

The problem turns out to be undecidable even for arbitrary linear context-free grammars.
It is worth mentioning that this problem is not equivalent to the problem of whether or not a given context-free grammar generates a regular language, which is known to be undecidable.

\begin{teor}\label{undecidable}
	Given a linear context-free grammar $G$, it is undecidable whether
	$dnreg_G(n)\in \mathcal O(1)$.
\end{teor}

If $L$ is a language generated by a context-free grammar such that every derivation of each word $w\in L$ of length $n$ needs a number of non-regular productions
at most linear in $n$, then the language is said to be ``at most linearly non-regular".
\begin{teor}
	$CF\subseteq DNREG(n)$. Every context-free language is at most linearly-non-regular.
\end{teor}

The next result gives an evaluation of the degree of non-regularity of unambiguous context-free grammar generating a non-regular language.

\begin{teor}\label{theorem5}
	Let $G$ be an unambiguous context-free grammar generating a non-regular
	language. Then $dnreg_G(n)\in \Omega(n)$.
\end{teor}

In \cite{jcss} one defines a complexity measure on pushdown automata which is related, to some extent, to the  pushdown space complexity of languages
introduced in \cite{Gabarro}. 

Let $\Gamma=(Q,V,U,\delta, q_0,Z_0, F)$ be a pushdown automaton with the set of states $Q$, the input alphabet $V$, the stack alphabet $U$,
the transition mapping $\delta$, the initial state $q_0$, the initial stack symbol $Z_0$ and the set of accepting states $F$. We say that a transition
$(s,\alpha)\in \delta (q,a,A)$, with $q,s\in Q$, $a\in V\cup\{\lambda\}$, $A\in U$, $\alpha\in U^*$, is a \emph{push move}, if $|\alpha|\ge 2$, it is a \emph{pop move} if $\alpha=\lambda$, and
it is a \emph{neutral move} if $\alpha\in U$. In \cite{jcss} one defines the {\it push complexity} of a language as the number of push moves
needed by a pushdown automaton to accept that language. Let $w\in V^*$ and 
$$C: (q_o,w,Z_0)\vdash^* (a,\lambda,\lambda)$$
be a computation in $\Gamma$ accepting the input word $w$ with empty stack, see, e.g., Chapter 6 in \cite{fla}. Then, the number of push moves in the computation $C$, is denoted by
$push_{\Gamma}(w,C)$. Furthermore, for every word $w\in V^*$ we define
\bean
push_{\Gamma} (w) &=& \left\{
\begin{array}{lll}
	min\{push_{\Gamma}(w,C)\mid $C$ \mbox{ is a computation accepting } w\}, \\
	\mbox{ if } w \mbox { is accepted by } \Gamma,\\
	0, \mbox{ if } w \mbox { is not accepted by } \Gamma.\\	
\end{array} \right.
\eean
We now define the function $push_{\Gamma}:\mathbb{N}\lra\mathbb{N}$ by
$$push_{\Gamma}(n)=max\{push_{\Gamma}(w)\mid |w|=n\}.$$
This function is called the push complexity of $\Gamma$.
Note that if a pushdown automaton has stack space complexity $f(n)$, its push complexity is a function $g(n)$ such that $f(n)\in\mathcal{O}(g(n))$.
As for the degree of non-regularity we set
\bean
PUSH_{\lambda}(f(n))&=&\{L\mid L=L(\Gamma) \mbox{ for some pushdown automaton } \Gamma \\
&& \mbox{ accepting with empty stack and }
push_{\Gamma}(n)\in \mathcal O(f(n))\}.
\eean

Analogously, we define 
\bean
PUSH_f(f(n))&=&\{L\mid L=L(\Gamma) \mbox{ for some pushdown automaton } \Gamma \\
&& \mbox{ accepting with final states and }
push_{\Gamma}(n)\in \mathcal O(f(n))\}.
\eean 

It is known how a pushdown automaton accepting with final states can be transformed into an equivalent one
accepting with empty stack (Theorem 5.1 in \cite{hopcroft}), and conversely (Theorem 5.2 in 
\cite{hopcroft}). By these constructions the equality $PUSH_{\lambda}(f(n))=PUSH_f(f(n))$.

The $push$ measure will turn out to be very useful for our further investigation. Indeed, we claim  that the two classes of languages $PUSH_{\lambda}(f(n))$ and $DNREG(f(n))$ are identical.

\begin{teor}
	Let $L$ be a deterministic context-free language that is not regular. If $L\in DNREG(f(n))$, then $f(n)\in\Omega(n)$.	
\end{teor}

\begin{teor}\label{radical}
	Both families $DNREG(\sqrt{n})$ and  $DNREG(\log n)$ contain
	non-regular languages.
\end{teor}

A very natural problem arises: Are there other sublinear functions $f$ such that
$DNREG(f)$ does contain non-regular languages?
The problem of finding other sublinear functions $f$ such that $DNREG(f)$ contains non-regular languages is of interest from a computational point of view as well. By the next theorem, functions
like $\log_p(n)$, for some $p\ge 2$, are of a special interest.

\begin{teor}\label{quadratic}
	Let $G$ be a context-free grammar in quasi Chomsky normal form generating a non-regular language such that $dnreg_G(n)\le f(n)$. Then $L(G)$ is recognizable in $\mathcal{O}(n\cdot p^{f(n)})$ time, where $p$ is the number of nonterminals of $G$.
\end{teor}

\section{The degree of extension of finite automata over groups}

Let $(M,\cdot,1)$ be a group under an operation denoted 
by $\cdot$ with the 
neutral element denoted by $1$.
An extended finite automaton (EFA shortly) $A$
over the group $(M,\cdot,1)$ is defined formally as follows.
$A=(Q,V,M,f,q_0,F)$, where
$Q,V,q_0,F$ have the same meaning as in a usual finite automaton, 
namely the set of states, the input alphabet, the initial state and the
set of final states, respectively, and $f:Q\times V\lra 
{\cal P}_f(Q\times M).$ This is actually the extension of finite automata with additive or multiplicative valences to an arbitrary group, see \cite{fernau} and the references therein.

This type of automaton can be viewed as a finite automaton having a register
in which any element of $M$ can be stored, let us call it ``group memory''. The relation
$(q,m)\in f(s,a),\ q,s\in Q,\ a\in V,\ m\in M$ means that the
automaton $A$ changes its current state $s$ into $q$, by reading the symbol
$a$ on the input tape, and stores $x\cdot m$ in the register, where $x$ is  the former content of the register.
The initial value stored in the register is $1$.

We shall use the notation
$$(q,aw,m)\models_A (s,w,mr)\mbox{ iff } (s,r)\in f(q,a)$$
for all $s,q\in Q\; a\in V,\; m,r\in M$.
The reflexive and transitive closure of the relation $\models_A$
is denoted by $\models_A^*$. Sometimes, the subscript identifying the automaton
will be omitted when it is self-understood.

The word $x\in V^*$ is accepted by the automaton $A$ if, and only if,
there is a final state $q$ such that $(q_0,x,1)\models^* (q,\eps,1)$.
In other words, a string is accepted if the automaton completely
reads the string and reaches a
final state with the content of the register being the neutral element of 
$M$.
The language accepted by an EFA $A$ over a group as above
is denoted by $L(A)$.

The following simple observation will be useful in what follows.
If $L$ is a language accepted by an EFA over some group $M$, there exists a finitely 
generated subgroup $N$ of $M$ such that $L$ is accepted by an EFA over $N$.
Indeed, since the EFA over some group has finitely many transitions, only finitely many
elements of the group can be associated with these transitions. 
Consequently, the register can only ever hold values in the
subgroup of the initial group generated by these elements, so it suffices to view the
automaton as an EFA over this subgroup.

It is clear that some words in the language accepted by an EFA over a group 
can be accepted by computations containing ``non-regular transitions'', that
is transitions that change the contents of the group memory. The use of these transitions can make EFA more powerful than nite
automata. A very simple example
is a finite automaton that accepts the language $\{a^nb^m\mid n,m\ge 1\}$. 
If we extend this automaton such that each transition reading an $a$ add the value $1$
to its register and each transition reading a $b$ subtracts $1$ from the register, the new
automaton is an EFA over the additive group of integers that accepts the non-regular language
$\{a^nb^n\mid n\ge 1\}$.
Consequently, EFA over groups are able to accept non-regular languages or even not context-free languages, see, e.g.,\cite{dassow}.
In the remainder of the present work we study ``how much'' group memory, defined as the number of
non-regular transitions, needs an EFA for accepting a non-regular language.

Given an EFA $A=(Q,V,M,f,q_0,F)$ over a group $(M,\cdot,1)$, $w\in L(A)$, and a computation 
$$C_A(w): (q_0,w_0,m_0)\models_A (q_1,w_1,m_1)\models_A (q_2,w_2,m_2)\models_A \dots \models_A (q_s,\eps,m_s),$$
for some $s\ge 1$, where $w_0=w$, $m_0=m_s=1$, we define 
the multiset $E(C_A(w))=\langle m_i^{-1}m_{i+1}\mid 0\le i\le s-1\rangle$. 
In words, $E(C_A(w))$ contains all the elements of $M$ used in the 
computation  $C_A(w)$, each element appearing in exactly the same number of copies as that of times that element was used during the computation.
Further on, let $N(C_A(w))$ be the integer defined by
$$N(C_A(w))=\sum_{x\in M, x\ne 1} E(C_A(w))(x).$$
We now define the 
{\it group memory complexity} of the computation of $A$ on the word $w$ by
$$gmc_A(w)= \left\{
\begin{array}{ll}
\min \{N(C_A(w))\mid C_A(w) \mbox{ is a computation of $A$ on $w$}\}\\
0, \mbox{ if } w\notin L(A).
\end{array}
\right.$$
In other words, the group memory complexity of a word with respect to an EFA over $M$ is computed by taking into consideration the “least non-regular computation” if there is one.
The group memory complexity of an EFA as above is a mapping from $\bbbn$ to $\bbbn$ defined by
$$gmc_A(n) =\max\{gmc_A(w) | |w|=n, w \in V^*\}.$$
As one can see, the most ``non-regular'' word of each length is considered.

Let $A$ be an arbitrary EFA over some group and $c$ be a positive integer; we define the language $L(A,\le c) = \{w \in L(A) \mid  gmc_A(w) \le  c\}$.
Clearly, if $gmc_A(n) \le c$ for any $n\ge 1$, then $L(A,\le c) =L(A)$ holds.
A natural question arises: Are there EFA accepting non-regular languages 
with a constant group memory complexity? We give a negative answer to the question through the following result:

\begin{teor}\label{constanta}
	Given an EFA $A$ and a positive integer $c$, the language $L(A,\le c)$ is regular.
\end{teor}

\begin{teor} Let $M$ be a group such that all its finitely generated 
	subgroups are finite. Then
	the language accepted by any EFA over $M$ is regular.
\end{teor} 

It is worth mentioning that the proof of Theorem \ref{constanta} is effective, that is 
a finite automaton recognizing $L(A,\le c)$ can effectively be constructed.
On the other hand, it is known that a pushdown automaton may be seen 
as an EFA over a free group \cite{dassow,Corson} or an EFA
over a polycyclic monoid \cite{chomsky,Shapiro}. Starting from these results we 
prove the next result.

\begin{teor}
	For every EFA $A$ over a free group or a polycyclic monoid and a positive integer $c$, the problem of whether or not $gmc_A(n)\le c$ is decidable. 
\end{teor}

Are there other classes of groups for which the question in the previous
statement is decidable? Yes, actually this happens
for every finitely generated abelian group. The reason is
a fundamental result in the group theory.

\begin{teor}
	Every finitely generated abelian group is the direct product of a finite
	number of cyclic groups.
\end{teor}
Consequently, the language accepted by an EFA over a finitely generated abelian group is either regular or is a language accepted by an EFA over
a group $\bbbz^k \times H$, where $k$ is a positive integer and $H$ is a finite abelian group. Moreover, $\bbbz^k$ is the additive group of
vectors of size $k$ with integer entries. We now make use of the next result (Theorem 7 in \cite{stiebe}):
\begin{teor}
	The language accepted by an EFA over an abelian group can be: (1) regular, (2) accepted by an EFA over by $\bbbz^k$, (3) accepted by an EFA over the multiplicative group of rationals.
\end{teor}

It follows that if a language $L$ accepted by an EFA over a finitely
abelian group is not regular, then there exists a positive integer $k$ such that $L$ is accepted by an EFA over $\bbbz^k$. We can now state

\begin{teor}
	For every EFA $A$ over a finitely generated abelian group and a positive integer $c$, the problem of whether or not $gmc_A(n)\le c$ is algorithmically decidable. 
\end{teor}

We now provide an EFA over an abelian group that
accept non-regular languages and has a sublinear group memory complexity,
namely a function in ${\mathcal O}(\sqrt{n})$.

\begin{lema}\label{ante_radical}
	There exists an EFA $A$ over $\bbbz\times \bbbz_2$ such that $L(A)$ is not regular and $gmc_A(n)\in {\mathcal O}(\sqrt{n})$.
\end{lema}

Inspired by the Goldstine language:

$$G=\{a^{n_1}ba^{n_2}b\dots a^{n_p}b\mid p\ge 1, n_i\ge 0, \mbox{ and } n_j\ne j \mbox{ for 
	some } j, 1 \le j \le p\},$$
we define the non-regular language
\bean
L&=& \{ba^{i_1}ba^{i_2}b\dots a^{i_k}bc^m\mid k\ge 1, i_1,i_2,\dots, i_k>0, \mbox{ and }\\
&& \mbox{ there exists } 1\le j\le k \mbox{ such that } i_j \ne j \mbox{ and } m=\left\{
\begin{array}{ll}
	j-i_j, \mbox{ if } j>i_j,\\
	1, \mbox{ if } j<i_j
\end{array}
\right.\}.
\eean
It can be routinely
proved that $L$ is not regular.

As it suffices to use Theorem 7 from \cite{stiebe} to simply replace the group 
$\bbbz\times \bbbz_2$ by $\bbbz$ in the statement of previous lemma, we can state:

\begin{teor}\label{radical2} Let $M$ be a group having at least one infinite cyclic subgroup.
	There exists an EFA $A$ over $M$ such that $L(A)$ is not regular and $gmc_A(n)\in {\mathcal O}(\sqrt{n})$.
\end{teor}

By using a similar idea to that used in the proof of Lemma \ref{ante_radical}
we prove the next result, where $\F_2$ is the free group of rank $2$. 

\begin{lema}\label{ante_log}
	There exists an EFA $A$ over the group $\F_2\times \bbbz_2$ such that $L(A)$ is not regular and $gmc_A(n)\in {\mathcal O}(\log n)$.
\end{lema}

As $\bbbz_2$ is a finite group, we state:

\begin{teor}\label{log}
	There exists an EFA $A$ over the group $\F_2$ such that $L(A)$ is not regular and $gmc_A(n)\in {\mathcal O}(\log n)$.
\end{teor}

We now give an example of a non-regular language such that any EFA over some group that accepts this language has a group memory complexity in $\Omega(n)$.

\begin{teor}\label{omega} 
	If $L(A)=\{a^nb^n\mid n\ge 1\}$, where $A$ is an EFA over some
	group, then $gmc_A(n)\in \Omega(n)$.
\end{teor}

Along these lines, two problems remain open here:
\begin{enumerate}
	\item Are there other abelian or non-abelian groups for which the aforementioned problem is decidable?
	\item Give a class of groups $\mathcal{M}$ such that for any group $M\in\mathcal{M}$ and
	an EFA $A$ over $M$ the problem of whether or not the group memory complexity of $A$ is finite
	is decidable/undecidable. 
\end{enumerate}

We have provided examples of EFA over some groups that
accept non-regular languages and have a sublinear group memory complexity,
namely a function in ${\mathcal O}(\sqrt{n})$ or ${\mathcal O}(\log {n})$.
Is it true that for any sublinear integer-valued function $f$, there is an EFA $A$ over some group $M$ such that
$L(A)$ is not regular and $gmc_A(n)\in\mathcal{O}(f(n))$?

Theorem \ref{omega} provides a non-regular language such that any EFA over some group
that accepts it has a linear group memory complexity.

It is worth mentioning that we have not considered here the deterministic variants of EFA over groups which will be investigated in another work.

\section{Jumping complexity of finite automata with translucent letters}

A {\it noneterministic finite automaton with translucent letters} (FATL) is a NFA
$M$ as above, such that the transition relation is defined in the following way.
First, we define the partial relation $\circlearrowright$ on the set of all
configurations of $M$: $(s,xay) \circlearrowright (p,xy)$ iff $p\in\delta(s,a)$,
and $\delta(s,b)$ is not defined for any $b\in alph(x)$, $s,p\in Q$,
$a,b\in V$, $x\in V^+$, $y\in V^*$.
We now write 
$$(p, x)\models_M (q, y), \mbox{ if either } (p, x)\ra (q, y) \mbox{ or } (p, x) \circlearrowright (q, y).$$
The subscript $M$ is omitted when it is understood from the context.

The language accepted by $M$ is defined by 
$$L(M)=\{x\in V^*\mid (q_0, x)\models^* (f,\eps), f\in F\}.$$
We want to stress that the automaton has been introduced in \cite{translucent}, with a slightly different definition. Actually, our definition is an FATL in the normal form in \cite{translucent} without a marker for the end of the input word. This automaton is also related to the {\it one way jumping automaton} introduced in \cite{jumping} with the difference that after each jump it returns to its previous position and does not make shift
of the jumped part to the end of the word. 

Let $M$ be an FATL; we consider $w\in L(M)$, and the accepting computation in $M$ on the input $w$:
$$C_M(w): (q_0, w)\models (q_1, w_1)\models (q_2, w_2)\models\dots \models (q_m,\eps),$$ with $q_i\in Q$, $1\le i\le m$, and $q_m\in F$. 
We define 
$$jc(C_M(w))=\{i\ge 1\mid (q_{i-1}, w_{i-1})  \circlearrowright (q_i, w_i)\}.$$
In words, $jc(C_M(w))$ contains all the jumping steps in the computation $C_M(w)$.

We now define the 
{\it jumping complexity} of the computation of $M$ on the word $w$ by
$$jc_M(w)= \left\{
\begin{array}{ll}
	\min \{card(jc(C_M(w)))\mid C_M(w) \mbox{ is a computation of $M$ on $w$}\}\\
	0, \mbox{ if } w\notin L(M).
\end{array}
\right.$$

In other words, the jumping complexity of a word with respect to $M$ is computed by taking into consideration the “least non-regular computation” if there is one. Equivalently, the jumping complexity of a word with respect to $M$ is the number of jumping steps of a computation with the minimal number
of jumping steps.

The jumping complexity of an automaton $M$ as above is a mapping from $\N$ to $\N$ defined by
$$jc_M(n) =\max\{jc_M(w) | |w|=n, w \in V^*\}.$$
As one can see, the most ``non-regular'' word of each length is considered.
It is clear that $jc_M(n)\le n$ for every jumping automaton $M$ as one letter is consumed in every step of a computation.

Let $f$ be a function from $\N$ to $\N$; we define the family of languages
$$LJC((f(n))=\{L\mid \exists \mbox{ FATL $M$ such that } L=L(M) \mbox{ and } jc_M(n)\in \mathcal{O}(f(n))\}.$$

Let $M$ be an arbitrary jumping automaton and $c$ be a positive integer; we define the language $L(M,\le c) = \{w \in L(M) \mid  jc_M(w) \le  c\}$.
Clearly, if $jc_M(n) \le c$ for any $n\ge 1$, then $L(M,\le c) =L(M)$ holds.
A natural question arises: Are there FATL accepting non-regular languages 
with a constant jumping complexity? We give a negative answer to the question through the following result:

\begin{lema}\label{constant2}
	Given an FATL $M$ and a positive integer $c$, the language $L(M,\le c)$ is regular.
\end{lema}

Consequently, we have 

\begin{teor}
	$JCL(1)$ equals the class of regular languages. 
\end{teor}
By Lemma \ref{constant2}, a sufficient condition for an FATL to accept a regular language is to be of constant jumping complexity. Is this condition necessary as well? Were this the case, the problem of deciding whether or
not the jumping complexity of a given FATL is constant would be undecidable.
Indeed, this decidability problem would be equivalent to decide whether or not
the language accepted by an FATL is a regular language, which is not decidable, see Proposition 8 in \cite{translucent}.
As it was expected, the condition is not necessary. It suffices to consider the FATL $M$ defined by the transition mapping:
$$\delta(q_0,b)=q_1, \delta(q_1,b)=q_1, \delta(q_1,c)=q_2,\delta(q_2,a)=q_3,$$
with the final state $q_3$. The language accepted by this automaton is 
$L=\{b^nab^mc\mid n+m\ge 1\}\cup\{b^nca\mid n\ge 1\}$, which is regular.
On the other hand, $jc_M(ab^nc)=n$, for any $n\ge 1$. 

However, the decidability status of the following related problem can be partially settled: {\it Given an FATL $M$ and a positive integer $c$, is it decidable whether or not $jc_M(n)\le c$ for all $n\ge 1$}?
By modifying the construction in the proof of Lemma \ref{constant2} we may infer:

\begin{lema}\label{constant3}
	Given a deterministic FATL $M$, there exists a deterministic FATL $M'$ such that $L(M')=\{w\in L(M)\mid jc_M(w)\ge 1\}$.
\end{lema}

This lemma is crucial for the next result.

\begin{teor}
	Given a deterministic FATL $M$ and a positive integer $c$, it is  algorithmically decidable whether or not the jumping complexity of $M$ 
	is bounded by $c$. 
\end{teor}

It is worth mentioning that even with this result, the decidability status
of the problem "Is the jumping complexity of a deterministic FATL finite?"
is still open.

Obviously, each FATL has a jumping complexity which is situated between the constant function and the identity function. In other words, $JCL(n)$ equals the class of all languages accepted by FATL. It remains to investigate what happens between these two extremes. First, we show that there are languages which require a jumping complexity in $\Omega(n)$.

\begin{teor}\label{omega2} 
	If $L(M)=\{w\mid |w|_a=|w|_b=n\ge 1\}$, where $M$ is an FATL, then $jc_M(n)\in \Omega(n)$.
\end{teor}

\begin{coro}
	$JCL(n)\setminus JCL(1)\ne\emptyset.$	
\end{coro}

\section{Final remarks}

There are still some attractive problems, in our opinion, that remained
unsolved here. One of the most intriguing is the existence of a lower bound for the degree of non-regularity for context-free languages which are not regular. As we have seen, there are context-free languages which are not regular having a sublinear degree of non-regularity. 
We strongly suspect a more general result: {\it $REG\subset DNREG(f)$, strict inclusion, for any function $f$ that is not a constant}. We mention a few other problems excepting the problem discussed before Theorem \ref{quadratic}. 
Given a context-free grammar $G$, is it decidable whether or not $G$ has the least degree of non-regularity among all grammars generating $L(G)$?

As far as the degree of extension of finite automata over groups is concerned, we have proved that given an EFA $A$ over a free group,
a polycyclic monoid, or a finitely generated abelian group and a constant $c$, one can
algorithmically decide whether or not the group memory complexity of $A$ is bounded by $c$.
Along these lines, two problems remain open here:
\begin{enumerate}
	\item Are there other abelian or non-abelian groups for which the aforementioned problem is decidable?
	\item Give a class of groups $\mathcal{M}$ such that for any group $M\in\mathcal{M}$ and
	an EFA $A$ over $M$ the problem of whether or not the group memory complexity of $A$ is finite
	is decidable/undecidable. 
\end{enumerate}

We have provided examples of EFA over some groups that
accept non-regular languages and have a sublinear group memory complexity,
namely a function in ${\mathcal O}(\sqrt{n})$ or ${\mathcal O}(\log {n})$.
Is it true that for any sublinear integer-valued function $f$, there is an EFA $A$ over some group $M$ such that
$L(A)$ is not regular and $gmc_A(n)\in\mathcal{O}(f(n))$?

Theorem \ref{omega} provides a non-regular language such that any EFA over some group
that accepts it has a linear group memory complexity.
It is worth mentioning that \cite{jlac} has not considered the deterministic variants of EFA over groups which is to be further investigated.

Some of the above problems remained open as well as regards the jumping complexity of finite automata with translucent letters.

Generally, this could be a measure for investigating the degree of extension of many mechanisms that extend a less expressive one like 
context-free grammars with regulated rewriting, extended various types of finite automata and tree automata over groups, jumping automata, automata with translucent letters, etc. A first step in this direction has already been done in \cite{robert}.

\nocite{*}
\bibliographystyle{eptcs}
\bibliography{afl_2023}

\begin{thebibliography}{10}
\providecommand{\bibitemdeclare}[2]{}
\providecommand{\surnamestart}{}
\providecommand{\surnameend}{}
\providecommand{\urlprefix}{Available at }
\providecommand{\url}[1]{\texttt{#1}}
\providecommand{\href}[2]{\texttt{#2}}
\providecommand{\urlalt}[2]{\href{#1}{#2}}
\providecommand{\doi}[1]{doi:\urlalt{https://doi.org/#1}{#1}}
\providecommand{\eprint}[1]{arXiv:\urlalt{https://arxiv.org/abs/#1}{#1}}
\providecommand{\bibinfo}[2]{#2}

\bibitemdeclare{article}{jlac}
\bibitem{jlac}
\bibinfo{author}{F.~\surnamestart Arroyo\surnameend},
  \bibinfo{author}{V.~\surnamestart Mitrana\surnameend},
  \bibinfo{author}{A.~\surnamestart P\u{a}un\surnameend},
  \bibinfo{author}{M.~\surnamestart P\u{a}un\surnameend} \&
  \bibinfo{author}{J.R.~\surnamestart S\'anchez Couso\surnameend}
  (\bibinfo{year}{2020}): \emph{\bibinfo{title}{On the group memory complexity
  of extended finite automata over groups}}.
\newblock {\slshape \bibinfo{journal}{J. Log. Algebraic Methods Program.}}
  \bibinfo{volume}{117}, p. \bibinfo{pages}{100605},
  \doi{10.1016/j.jlamp.2020.100605}.

\bibitemdeclare{article}{baker}
\bibitem{baker}
\bibinfo{author}{B.S. \surnamestart Baker\surnameend} (\bibinfo{year}{1974}):
  \emph{\bibinfo{title}{Non-context-free grammars generating context-free
  languages}}.
\newblock {\slshape \bibinfo{journal}{Information and Control}}
  \bibinfo{volume}{24}, pp. \bibinfo{pages}{231 -- 246},
  \doi{10.1016/S0019-9958(74)80038-0}.

\bibitemdeclare{book}{Balcazar}
\bibitem{Balcazar}
\bibinfo{author}{J.L. \surnamestart Balcazar\surnameend},
  \bibinfo{author}{J.~\surnamestart Diaz\surnameend} \&
  \bibinfo{author}{J.~\surnamestart Gabarr\'{o}\surnameend}
  (\bibinfo{year}{1995}): \emph{\bibinfo{title}{Structural Complexity}}.
\newblock \bibinfo{publisher}{Springer-Verlag, Berlin},
  \doi{10.1007/978-3-642-79235-9}.

\bibitemdeclare{article}{Boasson}
\bibitem{Boasson}
\bibinfo{author}{L.~\surnamestart Boasson\surnameend},
  \bibinfo{author}{B.~\surnamestart Courcelle\surnameend} \&
  \bibinfo{author}{M.~\surnamestart Nivat\surnameend} (\bibinfo{year}{1981}):
  \emph{\bibinfo{title}{The rational index, a complexity measure for
  languages}}.
\newblock {\slshape \bibinfo{journal}{SIAM J. Computing}} \bibinfo{volume}{10},
  pp. \bibinfo{pages}{284 -- 296}, \doi{10.1137/0210020}.

\bibitemdeclare{article}{Book}
\bibitem{Book}
\bibinfo{author}{R.V. \surnamestart Book\surnameend} (\bibinfo{year}{1972}):
  \emph{\bibinfo{title}{Terminal context in context-sensitive grammars}}.
\newblock {\slshape \bibinfo{journal}{SIAM J. Computing}} \bibinfo{volume}{1},
  pp. \bibinfo{pages}{20 -- 30}, \doi{10.1137/0201003}.

\bibitemdeclare{article}{jcss}
\bibitem{jcss}
\bibinfo{author}{H.~\surnamestart Bordihn\surnameend} \&
  \bibinfo{author}{V.~\surnamestart Mitrana\surnameend} (\bibinfo{year}{2020}):
  \emph{\bibinfo{title}{On the degrees of non-regularity and
  non-context-freeness}}.
\newblock {\slshape \bibinfo{journal}{J. Comput. Syst. Sci.}}
  \bibinfo{volume}{108}, pp. \bibinfo{pages}{104 -- 117},
  \doi{10.1016/j.jcss.2019.09.003}.

\bibitemdeclare{article}{Brainerd}
\bibitem{Brainerd}
\bibinfo{author}{B.~\surnamestart Brainerd\surnameend} (\bibinfo{year}{1968}):
  \emph{\bibinfo{title}{An analog of a theorem about context-free languages}}.
\newblock {\slshape \bibinfo{journal}{Information and Control}}
  \bibinfo{volume}{11}, pp. \bibinfo{pages}{561 -- 567},
  \doi{10.1016/S0019-9958(67)90771-1}.

\bibitemdeclare{inproceedings}{chomsky}
\bibitem{chomsky}
\bibinfo{author}{N.~\surnamestart Chomsky\surnameend} \& \bibinfo{author}{M.~P.
  \surnamestart Sch\"{u}tzenberger\surnameend} (\bibinfo{year}{1963}):
  \emph{\bibinfo{title}{The algebraic theory of context-free languages}}.
\newblock In: {\slshape \bibinfo{booktitle}{Studies in Logic and the
  Foundations of Mathematics}}, \bibinfo{publisher}{North-Holland, Amsterdam},
  pp. \bibinfo{pages}{118--161}, \doi{10.1016/S0049-237X(08)72023-8}.

\bibitemdeclare{article}{Corson}
\bibitem{Corson}
\bibinfo{author}{J.~M. \surnamestart Corson\surnameend} (\bibinfo{year}{2005}):
  \emph{\bibinfo{title}{Extended finite automata and word problems}}.
\newblock {\slshape \bibinfo{journal}{J. Algebra Comput.}}
  \bibinfo{volume}{15}, pp. \bibinfo{pages}{455 -- 466},
  \doi{10.1142/S0218196705002360}.

\bibitemdeclare{book}{Culik}
\bibitem{Culik}
\bibinfo{author}{K.~\surnamestart Culik\surnameend} \&
  \bibinfo{author}{H.~\surnamestart Maurer\surnameend} (\bibinfo{year}{1978}):
  \emph{\bibinfo{title}{On the Derivation of Trees}}.
\newblock \bibinfo{publisher}{TR Vol, 18, Institut f\"ur
  Informationsverarbeitung (Graz)}.

\bibitemdeclare{book}{regulated}
\bibitem{regulated}
\bibinfo{author}{J.~\surnamestart Dassow\surnameend} \& \bibinfo{author}{G.~\surnamestart P\u{a}un\surnameend} (\bibinfo{year}{1989}):
  \emph{\bibinfo{title}{Regulated Rewriting in Formal Language Theory}}.
\newblock \bibinfo{publisher}{Springer Berlin, Heidelberg},
  \doi{10.1007/978-3-642-74932-2}.

\bibitemdeclare{article}{dassow}
\bibitem{dassow}
\bibinfo{author}{J.~\surnamestart Dassow\surnameend} \&
  \bibinfo{author}{V.~\surnamestart Mitrana\surnameend} (\bibinfo{year}{2000}):
  \emph{\bibinfo{title}{Finite automata over free groups}}.
\newblock {\slshape \bibinfo{journal}{J. Algebra Comput.}}
  \bibinfo{volume}{10}, pp. \bibinfo{pages}{725 -- 737},
  \doi{10.1142/S0218196700000315}.

\bibitemdeclare{book}{Evey}
\bibitem{Evey}
\bibinfo{author}{R.~\surnamestart Evey\surnameend} (\bibinfo{year}{1963}):
  \emph{\bibinfo{title}{The Theory and Application of Pushdown Store
  Machines}}.
\newblock \bibinfo{publisher}{Doctoral Dissertation, Harvard University}.

\bibitemdeclare{article}{robert}
\bibitem{robert}
\bibinfo{author}{S.~Z. \surnamestart Fazekas\surnameend},
  \bibinfo{author}{R.~\surnamestart Merca\c{s}\surnameend} \&
  \bibinfo{author}{O.~\surnamestart Wu\surnameend} (\bibinfo{year}{2022}):
  \emph{\bibinfo{title}{Complexities for jumps and sweeps}}.
\newblock {\slshape \bibinfo{journal}{Journal of Automata, Languages and
  Combinatorics}} \bibinfo{volume}{27}, pp. \bibinfo{pages}{131--149},
  \doi{10.25596/jalc-2022-131}.

\bibitemdeclare{article}{fernau}
\bibitem{fernau}
\bibinfo{author}{H.~\surnamestart Fernau\surnameend} \&
  \bibinfo{author}{R.~\surnamestart Stiebe\surnameend} (\bibinfo{year}{2002}):
  \emph{\bibinfo{title}{Sequential grammars and automata with valences}}.
\newblock {\slshape \bibinfo{journal}{Theoret. Comput. Sci.}}
  \bibinfo{volume}{276}, pp. \bibinfo{pages}{377 -- 405},
  \doi{10.1016/S0304-3975(01)00282-1}.

\bibitemdeclare{inproceedings}{Gabarro2}
\bibitem{Gabarro2}
\bibinfo{author}{J.~\surnamestart Gabarro\surnameend} (\bibinfo{year}{1983}):
  \emph{\bibinfo{title}{Initial index:a new complexity function for
  languages}}.
\newblock In: {\slshape \bibinfo{booktitle}{International Colloquium on
  Automata, Languages, and Programming ICALP}}, \bibinfo{publisher}{{LNCS 154}
  Springer Verlag}, pp. \bibinfo{pages}{226--236}, \doi{10.1007/BFb0036911}.

\bibitemdeclare{inproceedings}{Gabarro}
\bibitem{Gabarro}
\bibinfo{author}{J.~\surnamestart Gabarro\surnameend} (\bibinfo{year}{1984}):
  \emph{\bibinfo{title}{Pushdown space complexity and related full-A.F.L.s}}.
\newblock In: {\slshape \bibinfo{booktitle}{Annual Symposium on Theoretical
  Aspects of Computer Science STACS}}, \bibinfo{publisher}{{LNCS 166} Springer
  Verlag}, pp. \bibinfo{pages}{250--259}, \doi{10.1007/3-540-12920-0{\_}23}.

\bibitemdeclare{article}{Shapiro}
\bibitem{Shapiro}
\bibinfo{author}{R.~H. \surnamestart Gilman\surnameend} \&
  \bibinfo{author}{M.~\surnamestart Shapiro\surnameend} (\bibinfo{year}{1998}):
  \emph{\bibinfo{title}{On groups whose word problem is solved by a nested
  stack automaton}}.
\newblock {\slshape \bibinfo{journal}{arXiv:math.GR/9812028}},
  \doi{10.48550/arXiv.math/9812028}.

\bibitemdeclare{article}{GG}
\bibitem{GG}
\bibinfo{author}{S.~\surnamestart Ginsburg\surnameend} \&
  \bibinfo{author}{S.~\surnamestart Greibach\surnameend}
  (\bibinfo{year}{1966}): \emph{\bibinfo{title}{Mappings which preserve
  context-sensitive languages}}.
\newblock {\slshape \bibinfo{journal}{Information and Control}}
  \bibinfo{volume}{9}, pp. \bibinfo{pages}{563 -- 582},
  \doi{10.1016/S0019-9958(66)80016-5}.

\bibitemdeclare{article}{Goldstine}
\bibitem{Goldstine}
\bibinfo{author}{J.~\surnamestart Goldstine\surnameend} (\bibinfo{year}{1972}):
  \emph{\bibinfo{title}{Substitution and bounded languages}}.
\newblock {\slshape \bibinfo{journal}{J. Comput. Syst. Sci.}}
  \bibinfo{volume}{6}, pp. \bibinfo{pages}{9 -- 29},
  \doi{10.1016/S0022-0000(72)80038-2}.

\bibitemdeclare{book}{Hibbard}
\bibitem{Hibbard}
\bibinfo{author}{T.~\surnamestart Hibbard\surnameend} (\bibinfo{year}{1966}):
  \emph{\bibinfo{title}{Scan Limited Automata and Context Limited Grammars}}.
\newblock \bibinfo{publisher}{Doctoral Dissertation, University of California
  at Los Angeles}.

\bibitemdeclare{book}{hopcroft}
\bibitem{hopcroft}
\bibinfo{author}{J.E. \surnamestart Hopcroft\surnameend} \&
  \bibinfo{author}{J.D. \surnamestart Ullman\surnameend}
  (\bibinfo{year}{1979}): \emph{\bibinfo{title}{Introduction to Automata
  Theory, Languages and Computation}}.
\newblock \bibinfo{publisher}{Addison-Wesley, Reading, Mass.}

\bibitemdeclare{article}{Kambites}
\bibitem{Kambites}
\bibinfo{author}{M.~\surnamestart Kambites\surnameend} (\bibinfo{year}{2006}):
  \emph{\bibinfo{title}{Word problems recognisable by deterministic blind
  monoid automata}}.
\newblock {\slshape \bibinfo{journal}{Theoret. Comput. Sci.}}
  \bibinfo{volume}{362}, pp. \bibinfo{pages}{232 -- 237},
  \doi{10.1016/j.tcs.2006.06.026}.

\bibitemdeclare{book}{Lyndon}
\bibitem{Lyndon}
\bibinfo{author}{R.~C. \surnamestart Lyndon\surnameend} \&
  \bibinfo{author}{P.~E. \surnamestart Schupp\surnameend}
  (\bibinfo{year}{1977}): \emph{\bibinfo{title}{Combinatorial Group Theory}}.
\newblock \bibinfo{publisher}{Springer-Verlag, Berlin},
  \doi{10.1007/978-3-642-61896-3}.

\bibitemdeclare{book}{fla}
\bibitem{fla}
\bibinfo{author}{C.~\surnamestart Mart\'{\i}n-Vide\surnameend},
  \bibinfo{author}{V.~\surnamestart Mitrana\surnameend} \& \bibinfo{author}{Gh.
  \surnamestart P\u{a}un\surnameend} (\bibinfo{year}{2003}):
  \emph{\bibinfo{title}{Formal Languages and Applications}}.
\newblock \bibinfo{publisher}{Springer Verlag},
  \doi{10.1007/978-3-540-39886-8}.

\bibitemdeclare{article}{Matthews1}
\bibitem{Matthews1}
\bibinfo{author}{G.~\surnamestart Matthews\surnameend} (\bibinfo{year}{1963}):
  \emph{\bibinfo{title}{Discontinuity and asymmetry in phrase structure
  grammars}}.
\newblock {\slshape \bibinfo{journal}{Information and Control}}
  \bibinfo{volume}{6}, pp. \bibinfo{pages}{137--146},
  \doi{10.1016/S0019-9958(63)90179-7}.

\bibitemdeclare{article}{Matthews2}
\bibitem{Matthews2}
\bibinfo{author}{G.~\surnamestart Matthews\surnameend} (\bibinfo{year}{1964}):
  \emph{\bibinfo{title}{A note on asymmetry in phrase structure grammars}}.
\newblock {\slshape \bibinfo{journal}{Information and Control}}
  \bibinfo{volume}{7}, pp. \bibinfo{pages}{360--365},
  \doi{10.1016/S0019-9958(64)90406-1}.

\bibitemdeclare{article}{Matthews3}
\bibitem{Matthews3}
\bibinfo{author}{G.~\surnamestart Matthews\surnameend} (\bibinfo{year}{1967}):
  \emph{\bibinfo{title}{Two-way languages}}.
\newblock {\slshape \bibinfo{journal}{Information and Control}}
  \bibinfo{volume}{10}, pp. \bibinfo{pages}{111--119},
  \doi{10.1016/S0019-9958(67)80001-9}.

\bibitemdeclare{article}{jumping}
\bibitem{jumping}
\bibinfo{author}{A.~\surnamestart Meduna\surnameend} \&
  \bibinfo{author}{P.~\surnamestart Zemek\surnameend} (\bibinfo{year}{2012}):
  \emph{\bibinfo{title}{Jumping finite automata}}.
\newblock {\slshape \bibinfo{journal}{Int. J. Found. Comput. Sci.}}
  \bibinfo{volume}{23}, pp. \bibinfo{pages}{1555--1578},
  \doi{10.1142/S0129054112500244}.

\bibitemdeclare{article}{stiebe}
\bibitem{stiebe}
\bibinfo{author}{V.~\surnamestart Mitrana\surnameend} \&
  \bibinfo{author}{R.~\surnamestart Stiebe\surnameend} (\bibinfo{year}{2001}):
  \emph{\bibinfo{title}{Extended finite automata over groups}}.
\newblock {\slshape \bibinfo{journal}{Discrete Appl. Math.}}
  \bibinfo{volume}{108}, pp. \bibinfo{pages}{287 -- 300},
  \doi{10.1016/S0166-218X(00)00200-6}.

\bibitemdeclare{inproceedings}{translucent}
\bibitem{translucent}
\bibinfo{author}{B.~\surnamestart Nagy\surnameend} \& \bibinfo{author}{L.~\surnamestart Kov\'acs\surnameend} (\bibinfo{year}{2014}):
  \emph{\bibinfo{title}{Finite automata with translucent letters applied in
  natural and formal language theory}}.
\newblock In: {\slshape \bibinfo{booktitle}{Transactions on Computational
  Collective Intelligence}}, \bibinfo{publisher}{{LNCS 8790} Springer Verlag},
  pp. \bibinfo{pages}{107--127}, \doi{10.1007/978-3-662-44994-3{\_}6}.

\bibitemdeclare{book}{Rotman}
\bibitem{Rotman}
\bibinfo{author}{J.~J. \surnamestart Rotman\surnameend} (\bibinfo{year}{1995}):
  \emph{\bibinfo{title}{An Introduction to the Theory of Groups}}.
\newblock \bibinfo{publisher}{Springer-Verlag, Berlin},
  \doi{10.1007/978-1-4612-4176-8}.

\bibitemdeclare{book}{handbook}
\bibitem{handbook}
\bibinfo{author}{G.~\surnamestart Rozenberg\surnameend} \&
  \bibinfo{author}{A.~\surnamestart Salomaa\surnameend} (\bibinfo{year}{1997}):
  \emph{\bibinfo{title}{Handbook of Formal Languages}}.
\newblock \bibinfo{publisher}{Springer Verlag},
  \doi{10.1007/978-3-642-59136-5}.

\end{thebibliography}
\end{document}